# Large Language Models Predict Human Well-being—But Not Equally Everywhere

Pat Pataranutaporn, *Massachusetts Institute of Technology, Cambridge, MA, USA*

Nattavudh Powdthavee, *Nanyang Technological University, Singapore*

Chayapatr Achiwaranguprok, *Massachusetts Institute of Technology, Cambridge, MA, USA*

Pattie Maes, *Massachusetts Institute of Technology, Cambridge, MA, USA*

**Abstract**

Subjective well-being is a key metric in economic, medical, and policy decision-making. As artificial intelligence provides scalable tools for modelling human outcomes, it is crucial to evaluate whether large language models (LLMs) can accurately predict well-being across diverse global populations. We evaluate four leading LLMs using data from 64,000 individuals in 64 countries. While LLMs capture broad correlates such as income and health, their predictive accuracy decreases in countries underrepresented in the training data, highlighting systematic biases rooted in global digital and economic inequality. A pre-registered experiment demonstrates that LLMs rely on surface-level linguistic similarity rather than conceptual understanding, leading to systematic misestimations in unfamiliar or resource-limited settings. Injecting findings from underrepresented contexts substantially enhances performance, but a significant gap remains. These results highlight both the promise and limitations of LLMs in predicting global well-being, underscoring the importance of robust validation prior to their implementation across these areas.

**Introduction**

Subjective well-being — particularly overall life satisfaction — is increasingly recognised as a core measure of human flourishing and a critical input into economic, medical, and policy decision-making.[1-4] While many countries now collect well-being data, few have systematically integrated these metrics into policymaking.[5-7] This challenge is especially acute in lower-resource settings, where high-quality, nationally representative data are costly, logistically difficult, or entirely unavailable.[8] Even in data-rich contexts, traditional surveys are infrequent and vulnerable to respondent fatigue[9] and social desirability bias,[10] limiting their ability to capture dynamic changes in well-being.

Recent advances in large language models (LLMs) offer a promising, low-cost alternative.[11-12] Trained on massive corpora of human text, LLMs can infer subjective well-being from indirect signals — such as free-text reflections, demographic profiles, and contextual indicators — potentially circumventing the need for formal questionnaires. This opens avenues for real-time, high-resolution monitoring and modeling of human well-being, especially in contexts where traditional data collection is impractical. Yet it remains unclear whether LLMs can produce accurate and equitable estimates of life satisfaction across diverse global contexts, or if they simply generalise from the experiences of more digitally visible populations. While LLMs have excelled in cognitive and linguistic benchmarks[13-15], their effectiveness in modeling subjective, culturally embedded constructs like life satisfaction remains largely untested.

In this study, we assess the performance of four prominent LLMs—GPT, Claude, LLaMA, and Gemma—using a balanced and harmonised dataset of 64,000 individuals from 64 countries. We benchmark model predictions against both self-reported life satisfaction and conventional statistical techniques (OLS and Lasso regression). While LLMs broadly capture well-known correlates such as income, education, health, and perceived freedom, they consistently underperform traditional models and systematically flatten meaningful cross-country variation.

We identify two key sources of error: training data bias from high-income contexts[16-20] and semantic overgeneralisation. While prior work has documented algorithmic bias in identity-sensitive domains, such as race and gender, our findings show how these limitations extend to the global level—where cultural, economic, and linguistic diversity is high but underrepresented in training corpora. First, because LLMs are primarily trained on text from wealthier, digitally well-represented countries, our experiment reveals their predictions disproportionately reflect the values and priorities of these contexts—resulting in systematic misestimation of life satisfaction in low-resource and digitally marginalised countries. Second, LLMs often rely on linguistic proximity: they overweight conceptually salient but empirically weak predictors (e.g., education, democracy) and underweight stronger but subtler predictors of well-being in resource-poor countries (e.g., income, perceived freedom). This semantic overreach is especially problematic in cross-cultural applications, where language cues do not always align with lived experiences.

These results challenge the optimistic view that LLMs can democratise behavioural insight. They shift the debate from *who* is represented in training data to *how* meaning is inferred, thereby highlighting the risk that AI predictions may obscure inequality rather than reveal it. Our findings underscore the need for robust validation frameworks before deploying LLMs in global health, development, or policy contexts.

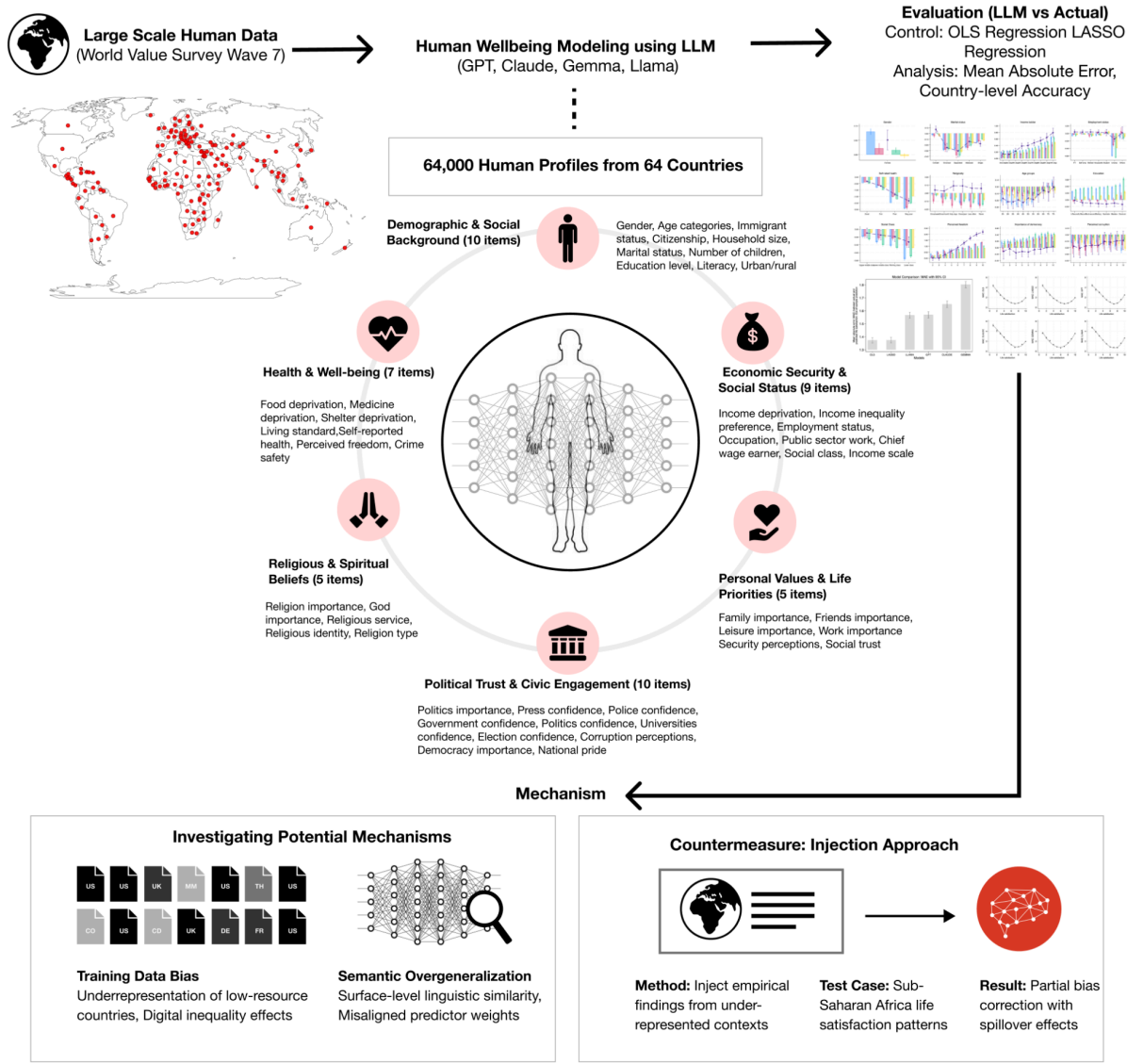

**Fig 1:** The experimental framework evaluates large language models' ability to predict individual life satisfaction across 64 countries, benchmarking LLMs' performance against traditional statistical approaches using World Values Survey data from 64,000 respondents. The study also examines semantic overgeneralization mechanisms and tests targeted injection methods to determine whether systematic biases in underrepresented regions can be mitigated.

## Results

### Benchmarking LLM Performance Against Statistical Models

In our pre-registered experiments, we evaluated the ability of four large language models — GPT-4o mini, Claude 3.5 Haiku, LLaMA 3.3 70B, and Gemma 3 27B — to predict individual life satisfaction across 64 countries using observational data from the World Values Survey (N = 64,000). For benchmarking, we compared out-of-sample LLM predictions to traditional statistical models, including ordinary least squares (OLS) and Lasso regressions, and to participants' self-reported life satisfaction scores. Although LLMs captured broad patterns linking income, education, health, and perceived freedom to life satisfaction, their overall predictive accuracy lagged behind that of traditional models. As shown in Fig. 2 (Panel A), ordinary least squares (OLS) and Lasso achieved the lowest mean absolute errors (MAEs), both at 1.37 (95% CI: 1.35–1.39). In contrast, all LLMs—LLaMA, GPT, Claude, and Gemma—performed substantially worse, with MAEs ranging from 1.57 (LLAMA and GPT; 95% CI: 1.54–1.59) to 1.80 (Gemma; 95% CI: 1.78–1.83), reflecting an increase in average prediction error of up to 0.43 points, or roughly 31%, relative to OLS. Given that a 0.4-point shift in life satisfaction is comparable in magnitude to the effects of major life events such as unemployment or marriage[21], these discrepancies are both statistically and practically meaningful.

Further analyses (Fig.2, Panels B & C) reveal that absolute prediction errors are not evenly distributed across the life satisfaction scale. All models exhibited a pronounced U-shaped pattern, with the largest errors occurring among individuals reporting low life satisfaction (≤4) and a secondary increase among those reporting the highest scores (≈10). Absolute error rates were lowest for individuals in the mid-to-high range (6–8). This pattern was particularly pronounced for LLMs, suggesting that these models struggle most to predict life satisfaction among individuals at the margins of the distribution—especially the least satisfied.

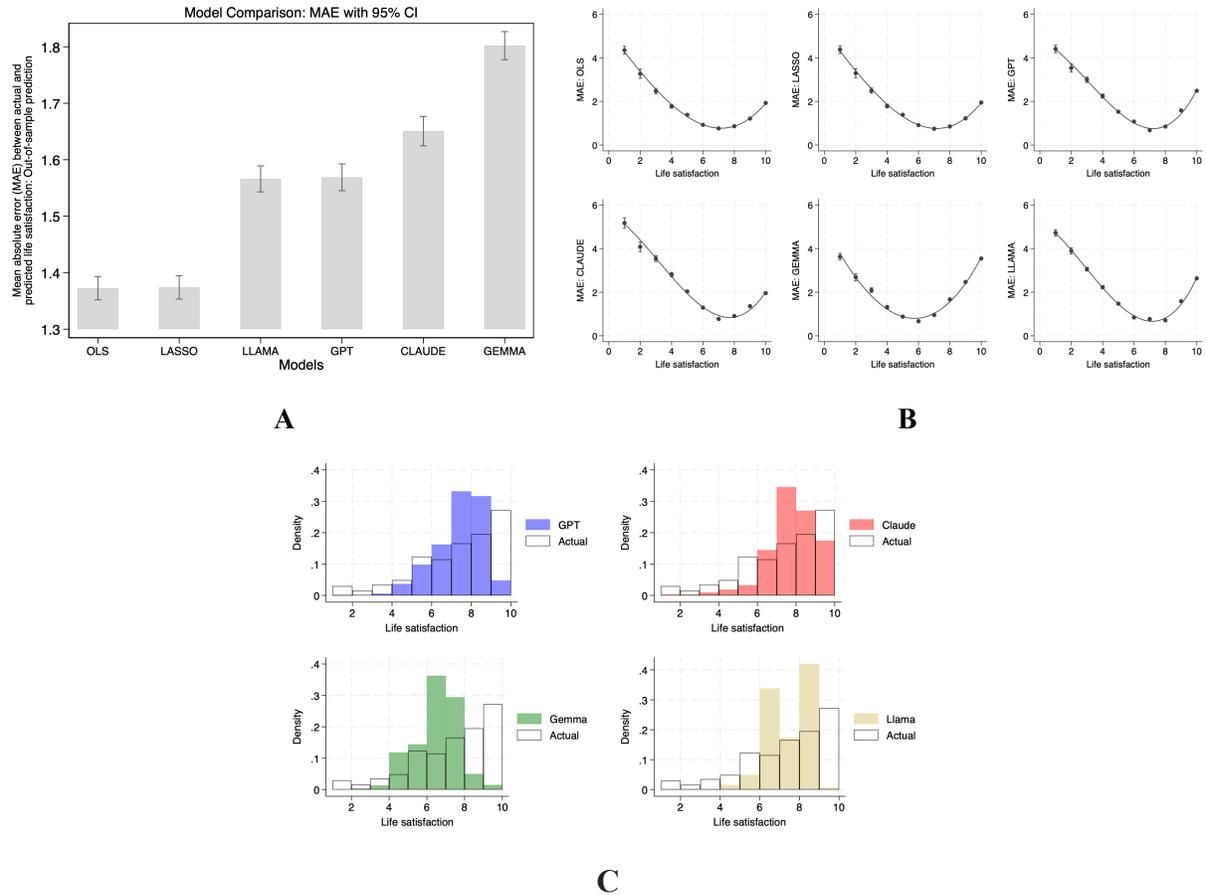

**Fig 2:** Panel A–Differences in out-of-sample performance between OLS, Lasso, and large language models. The standard error bars represent the 95% confidence intervals. Panel B–Comparison of mean absolute error in out-of-sample predictions (20% test sample) across regression and large language models, binned by actual life satisfaction. Panel C–Distribution of (i) actual life satisfaction scores and (ii) LLM-predicted life satisfaction scores.

**Overestimation and Underestimation of Life Satisfaction Predictors**

To understand the sources of LLMs' misestimation of extreme values of life satisfaction, we next investigate which individual-level predictors they misrepresent. We compared the determinants of actual life satisfaction in survey data with those implied by LLM-generated estimates. Figure 3 presents the relative influence of socioeconomic and psychological factors, contrasting OLS coefficients from observed data (dotted lines) with those produced by four leading LLMs (coloured bars).

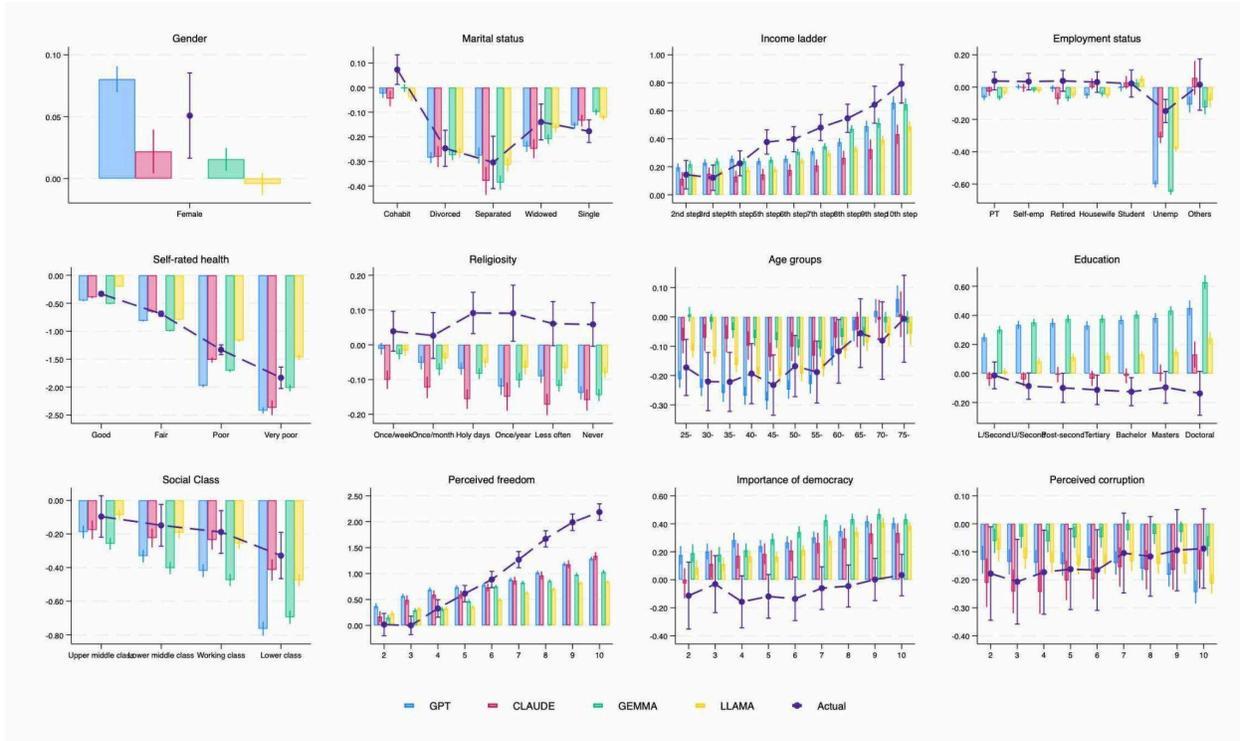

**Fig 3:** Coefficient plot of socio-economic and attitudinal predictors of actual and LLM-predicted life satisfaction. Reference groups are: male (gender), married (marital status), first step (income ladder), full-time employed (employment status), very good (self-rated health), more than once a week (religiosity), younger than 25 (age groups), no formal education (education), upper class (social class), 1.little perceived freedom (perceived freedom), 1.democracy is not important (importance of democracy), and 1.zero corruption (perceived corruption). 95% confidence intervals are displayed.

Several clear divergences emerge. For example, perceived freedom—strongly associated with higher life satisfaction in the observed data—is consistently underweighted across all four models, indicating that LLMs fail to fully reflect the psychological costs of low autonomy. A similar pattern appears for income: the steep well-being gradient from '2$^{nd}$ step' to '10$^{th}$ step' in the income distribution is notably compressed in all four models.

In contrast, LLMs tend to assign greater weight to certain attributes than is supported by real-world data. Education is a notable example: although its direct association with life satisfaction—once factors such as income, marital status, and occupational status are accounted for—is modest or even flat, particularly at higher levels as well-documented in the literature[22-23], three out of four models (with the exception of Claude) predict a strong, upward-sloping relationship. A similar overestimation appears for the importance of democracy, which is given greater weight in the models than supported by actual survey responses. For negative predictors such as unemployment and poor health, the pattern is reversed but no less pronounced: while both are strongly associated with lower well-being in the data, LLMs consistently predict even larger penalties. Together, these discrepancies reveal a systematic misalignment between model-generated inferences and the true explanatory power of these variables.

To evaluate these discrepancies more systematically, we regressed the absolute prediction error against individual characteristics (Extended Data Fig. 1a). The findings show significant error gradients across the entire spectrum of income, self-reported health, and perceived freedom, indicating that model misestimation escalates toward either extreme of these variables. These trends imply that existing LLMs have difficulty aligning with empirically verified patterns of life satisfaction.

**LLMs Attenuate Country-Specific Variation**

Having examined individual-level predictors, we next assessed whether LLMs capture systematic country-level differences in life satisfaction that remain unexplained after accounting for observed factors such as income, health, and employment—potentially reflecting cultural, linguistic, or institutional variation. Figure 4 presents the estimated country fixed effects for both actual and LLM-predicted outcomes, adjusted for individual-level characteristics such as income, health, education, and employment. In the observed data, country effects varied widely—from –2.1 to +1.0—highlighting substantial cross-national heterogeneity beyond compositional factors. Nations such as Mexico, Colombia, and Peru reported significantly above-average well-being, consistent with prior findings from Latin America.[24] Conversely, countries like Ethiopia, Iraq, and Zimbabwe recorded the lowest adjusted scores.

All four LLMs substantially understated this variation. Fixed effects clustered tightly around zero, typically between –0.5 and +0.5, with Llama producing the narrowest range (±0.2). Extreme cases were notably misrepresented: Mexico, Colombia, and Peru were assigned near-zero or even negative effects, while Ethiopia, Iraq, and Zimbabwe appeared only modestly below average. Although Claude and Gemma yielded slightly more dispersion than GPT or Llama, none reproduced the magnitude or structure of cross-country differences observed in the actual data. Crucially, this flattening occurred despite explicitly including country of residence in the input, suggesting that current LLMs struggled to integrate national context in a meaningful way. Further robustness checks (Extended Data Fig. 2a) reinforce earlier findings, revealing substantial cross-national variation in absolute prediction errors—ranging from the Netherlands to Ethiopia for GPT and Llama, Malaysia to Ethiopia for Claude, and Iraq to Mexico for Gemma.

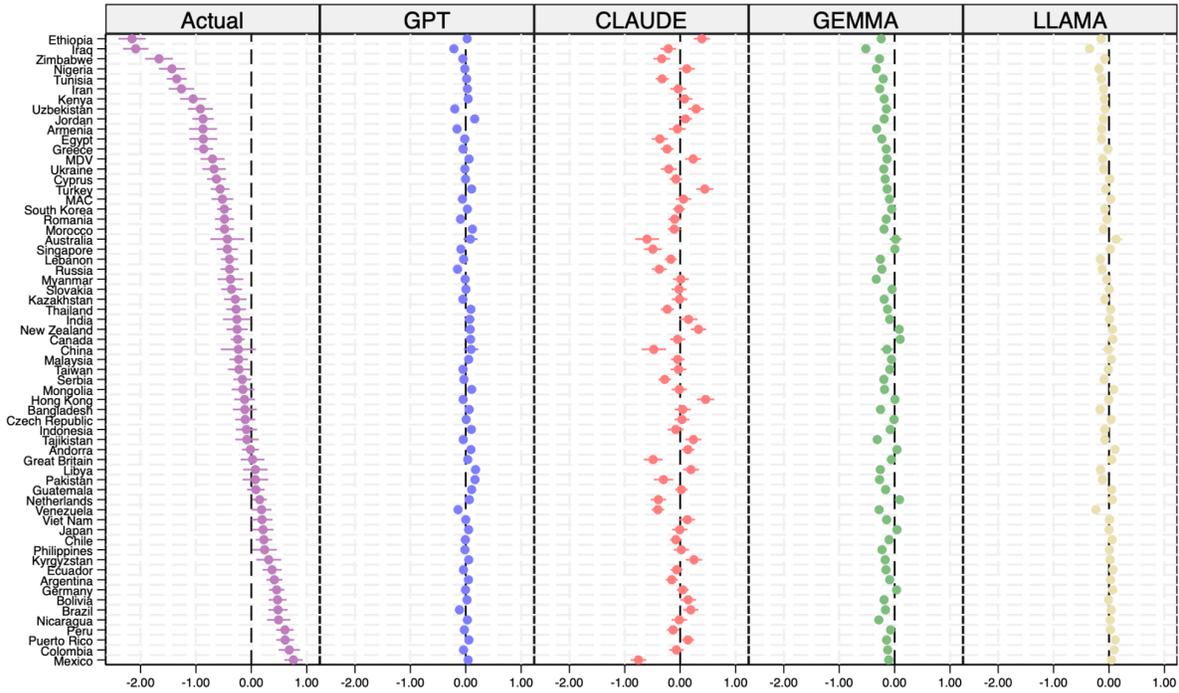

**Fig 4:** Coefficient plot of country fixed effects on actual and LLM-predicted life satisfaction. Reference group: USA. 95% confidence intervals are displayed

Taken together, these findings show that LLMs misestimate the drivers of life satisfaction, overstating the effects of variables like poor health, unemployment, and education, while compressing gradients for income and perceived freedom. They also fail to capture the wide cross-national variation observed in survey data, despite having access to country information. These distortions suggest that current models reflect broad associations but miss the underlying structure of well-being.

**Training data bias and semantic generalisation as sources of misestimation**

These systematic misestimations raise a critical question: why do LLMs, despite having access to detailed individual and contextual information, misestimate certain empirical structures of life satisfaction? To explore this, we next consider two interrelated sources of error—**training data biases** and **semantic generalisation**—that may underlie the observed patterns.

**Training data biases.** Because LLMs are trained on internet-scale corpora dominated by content from high-income, English-speaking, and digitally connected regions[25], their representations of life satisfaction may reflect the priorities and associations of those contexts. This can result in systematic mispredictions in underrepresented regions, particularly in how individual-level predictors relate to well-being. To test this, we modelled the absolute prediction error as a function of standard individual-level variables (e.g., income, employment, health) and averaged the resulting fitted values by country. These country-level averages reflect the extent to which prediction errors can be systematically explained by the observed individual predictors. As shown in Figure 5 (Panel A), these values were negatively correlated with HDI, log GDP per capita, and internet usage, and positively correlated with the Gini index, thus indicating that all four LLMs more accurately capture the structure of individual-level predictors in high-resource, digitally connected countries.

In addition to these predictor-level misalignments, we also assessed whether LLMs exhibit elevated "unexplained" errors at the country level, which represent country-specific residuals in the life satisfaction equation (e.g., culture, language, social norms), even after accounting for individual characteristics. Specifically, we regressed the country fixed effects from Figure 4 on macro-level indicators. As shown in Figure 5 (Panel B), country-specific residuals for GPT, Gemma, and Llama increased sharply in countries with lower HDI, GDP, and internet access. The Gini index showed a weaker but positive association, suggesting greater modelling difficulty in more unequal societies. Claude's fixed effects, by contrast, were largely uncorrelated with these indicators, pointing to a flatter, less context-sensitive error profile.

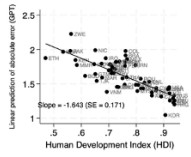
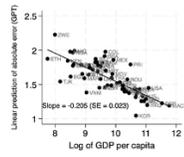
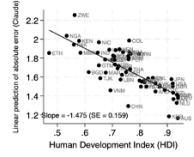
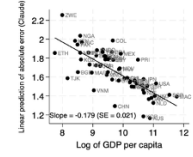
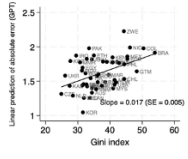
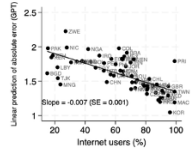
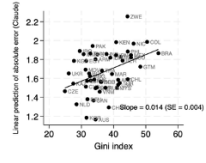
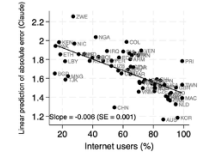

a) GPT

b) Claude

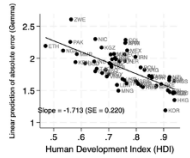
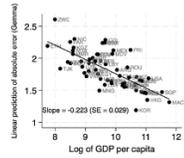
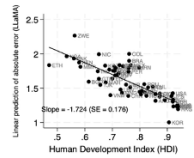
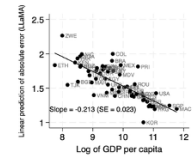
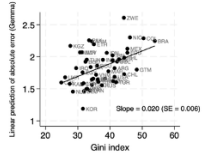
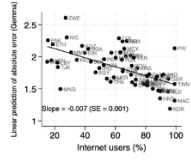
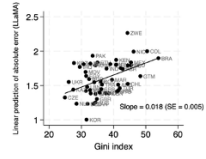
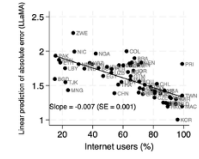

c) Gemma

d) Llama

**A**

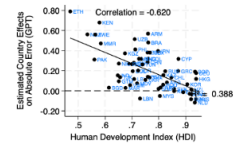
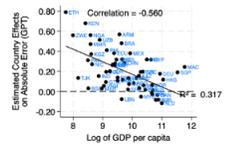
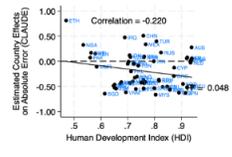
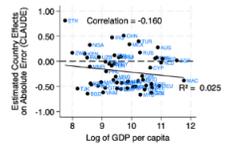
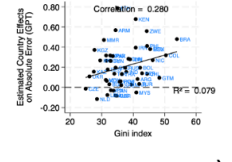
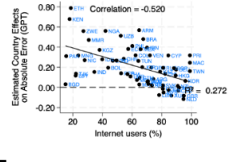
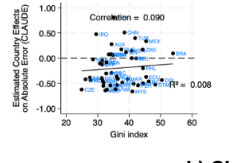
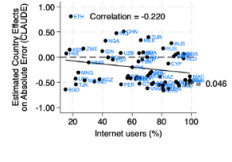

a) GPT

b) Claude

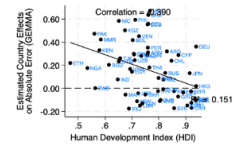
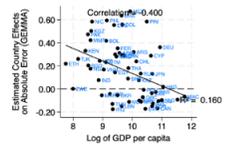
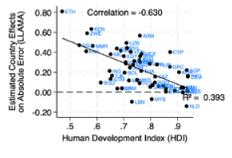
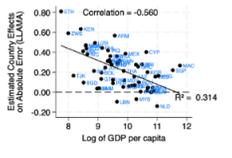
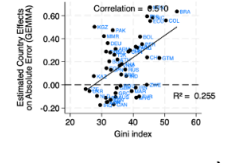
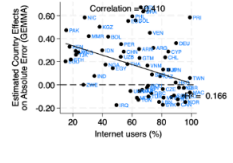
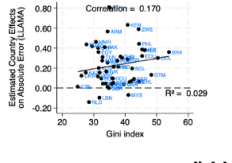
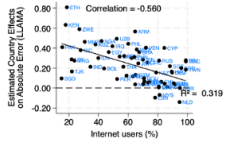

c) Gemma

d) Llama

**B**

**Figure 4**: Panel A–Two-way scatter plots showing country-level averages of fitted values from regressions of absolute prediction error on individual-level predictors, plotted against macroeconomic indicators. Panel B–Two-way scatter plots showing estimated country fixed effects from regressions of absolute prediction error on individual-level predictors, plotted against macroeconomic indicators. The macroeconomic indicators include: Human Development Index (HDI), log GDP per capita, Gini index (income inequality), and internet usage (% of population). Higher values indicate greater systematic error explainable by observed predictors.

To illustrate systematic misestimation in lower-resource settings, Figure 3a in the appendix illustrates systematic misestimation by LLMs across development contexts. In lower-HDI countries, the observed relationship between income and life satisfaction is markedly steeper than in higher-HDI countries, reflecting the greater marginal value of income. However, all four LLMs predict nearly identical income–satisfaction gradients across both groups—underestimating the effect in low-HDI settings and overestimating it in high-HDI ones. This suggests that LLMs capture a homogenised income–well-being relationship that overlooks critical cross-context variation.

Together, these results show that LLMs' predictive accuracy declines most in underrepresented or digitally marginalised countries—precisely those least visible in internet-scale training data—offering a plausible explanation for both their systematic misestimation of individual-level predictors and the muted country fixed effects observed in these settings.

**Semantic generalisation**. A key explanation for LLMs' prediction errors is their tendency to rely on linguistic associations rather than abstract reasoning, which leads to surface-level extrapolations that diverge from empirical relationships. For example, public discourse often frames education as a marker of success, potentially prompting models to overestimate its effect—even when survey data show weak associations once income, health, and marital status are controlled for.[23] Similarly, the psychological cost of unemployment, widely emphasised in Western narratives, is often overgeneralised. These patterns suggest that LLMs may generalise along semantic lines, guided by proximity in language rather than by deeper conceptual structures. Just as "happiness" and "life satisfaction" differ in psychological structure despite overlapping in language[26], LLMs may treat conceptually distinct prompts as interchangeable if they are semantically adjacent, thus raising concerns about their reliability in novel or policy-relevant applications where precise distinctions matter.

To test this, we designed a randomised experiment with four synthetic interventions. Each introduced a pair of fictional variables, one positively and one negatively associated with life satisfaction. These variables were entirely fabricated (e.g., "listening to unicorn voices" vs. "having a dinosaur companion") and connected through five intermediate prompts that formed a smooth semantic gradient. Only the endpoints (conditions 1 and 7) were seen by the models during context injection; the intermediate prompts (conditions 2–6) were novel, allowing us to test whether LLMs extrapolate from injected associations to unseen but semantically related inputs.

Every condition was evaluated using 400 synthetic prompts, resulting in 2,800 responses per model for each intervention. Figure 6 outlines the results for four prominent LLMs: GPT, Claude, Gemma, and Llama. Three of the models—GPT, Gemma, and LLaMA—demonstrated clear semantic interpolation; their predictions for life satisfaction changed smoothly across the spectrum, even though the intermediate prompts were not previously seen. Claude also displayed a directional shift, but this was characterised by a flatter gradient and revealed no statistically significant differences from the control group in any of the individual conditions.

Reversing the mapping (i.e., switching which endpoint was associated with higher life satisfaction) inverted the gradient in all models, suggesting that predictions tracked injected semantics rather than inherent properties of the prompts. Placebo controls, trained on semantically irrelevant but stylistically similar text, showed no such gradient, reinforcing that the effect stemmed from semantic similarity, not generic exposure.

The strength and shape of these interpolation patterns differed notably. GPT and Gemma displayed sharp, consistent gradients, indicative of strong generalisation. LLaMA's responses were shallower and more erratic across conditions. Claude's behaviour was the most restrained: while its predictions followed the injected direction, deviations from control and placebo were minimal and statistically insignificantly different from zero. This suggests a conservative generalisation strategy that may limit semantic overreach.

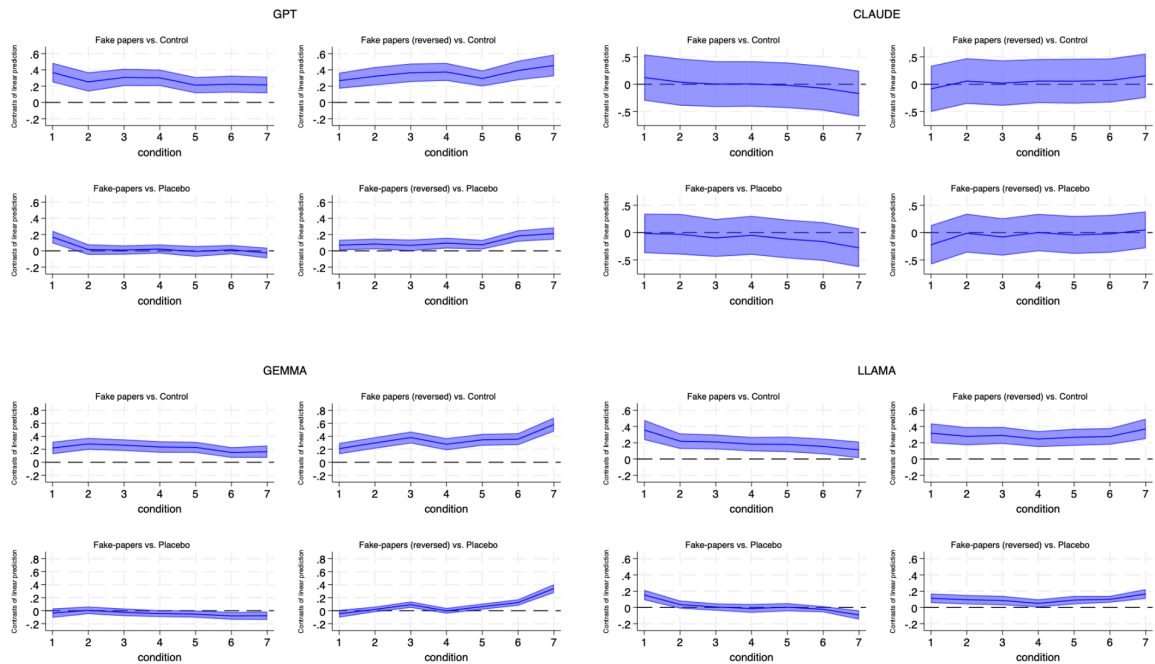

**Fig 6**: Treatment effects on predicted life satisfaction across semantic conditions and LLMs. Each panel shows estimated differences in predicted life satisfaction between treated and comparison models across seven semantic levels. Top row: models with fabricated research papers injection vs. the baseline model (left: original framing; right: reversed). Bottom row: same models vs. a placebo model with an injection of unrelated synthetic content. Only conditions 1 (positive) and 7 (negative) were seen during context injection; conditions 2–6 represent unseen, semantically interpolated prompts. Shaded areas show 95% confidence intervals. All regressions control for intervention fixed effects. Systematic deviations across intermediate conditions indicate semantic generalisation beyond the trained endpoints.

These findings together show that LLMs often consider linguistic similarity as conceptual equivalence. While this may be efficient in familiar domains, it poses risks in novel or policy-sensitive contexts where linguistic cues can be misleading. Models that generalise more aggressively—like GPT and Gemma—are more responsive to injected signals but also more susceptible to overextension. In contrast, more conservative models—like Claude—may preserve conceptual fidelity but at the cost of reduced responsiveness. These dynamics highlight a key trade-off in model behaviour: the balance between the strength of generalisation and the maintenance of semantic boundaries. They also suggest that the effectiveness of injection varies across models and contexts—that is, the injection effect is heterogeneous and may depend on how salient or conceptually anchored the injected information is. In the following section, we focus on Claude as a test case to assess whether selective injection can be used to *correct specific* mispredictions, leveraging its more cautious generalisation to minimise unintended spillover effects.

**Can targeted injection correct semantic overgeneralisation and training data bias?**

One of the most pronounced errors made by LLMs is the systematic overestimation of life satisfaction in Sub-Saharan countries—particularly Ethiopia, Zimbabwe, Nigeria, and Kenya—relative to self-reported data. To evaluate whether this bias could be reduced, we conducted an intervention in which a general empirical finding was injected into the model: that *"average life satisfaction in Sub-Saharan countries is typically lower than in most other regions."* Building on earlier results, we use Claude as a test case, given its tendency towards more cautious generalisation and reduced risk of unintended spillover. We then examined how this targeted injection affected predictions across 14 countries (N = 14,000), focusing both on the directly relevant Sub-Saharan countries and on less directly targeted regions. These included five Latin American countries—Argentina, Brazil, Chile, Colombia, and Mexico—where LLMs tend to underpredict life satisfaction, and five others—Canada, Germany, Japan, Myanmar, and the United States—which exhibit relatively similar average levels of life satisfaction. This broader sample allows us to assess not only the intended effects of the injection but also its potential spillover across regions that were not explicitly referenced in the injected content.

We highlight two key results, presented in Figure 7**.** Panel A displays coefficient plots comparing country-specific effects on actual self-reported life satisfaction with Claude's predictions under both non-injected and injected conditions. Country fixed effects are estimated relative to the United States, which serves as the reference category. Recall that these fixed effects represent systematic country-level differences in life satisfaction that remain unexplained after controlling for individual-level characteristics such as income, health, and employment, potentially reflecting cultural, linguistic, or institutional variation. In the absence of injection, Claude predicted life satisfaction in Sub-Saharan countries to be comparable to—or even higher than—that of the United States, despite empirical data indicating considerably lower average scores. Following the injection, however, all four Sub-Saharan countries now have significantly lower life satisfaction than the reference group, bringing the model's outputs closer to aligning with observed data. This suggests that the injection was successful in shifting the level of predicted well-being in the intended direction. Nonetheless, the gap persists, indicating only partial correction of the original bias.

Panel B presents predictions of absolute error by region and treatment condition, based on linear models that incorporate individual-level characteristics such as gender, age, income, and health. Baseline errors are highest in Sub-Saharan countries, with the injection resulting in a substantial reduction in this group. In contrast, Latin American and other countries—neither of which was directly referenced in the injected content—serve as a test of potential spillover effects. While baseline errors in these regions are smaller, we observe noticeable changes in predicted life satisfaction following injection, suggesting that even targeted factual input can influence model behaviour beyond its intended scope.

Changes in the linear predictions of absolute error suggest that the injection may have modified how Claude internally weighs individual characteristics when estimating life satisfaction. Although the injected message related only to average regional differences—namely, that life satisfaction tends to be lower in Sub-Saharan Africa—it may have inadvertently altered the model's internal feature mappings, with early evidence of variation in slope estimates between injected and non-injected conditions. These findings underscore both the potential and the limitations of selective injection: while it can enhance calibration in underrepresented contexts, it may also introduce unanticipated changes in model generalisation, emphasising the need for careful design, testing, and model-specific evaluation. For a breakdown of individual predictors of actual and model-predicted life satisfaction by treatment and region, see Figs. 5a–7a in the Appendix.

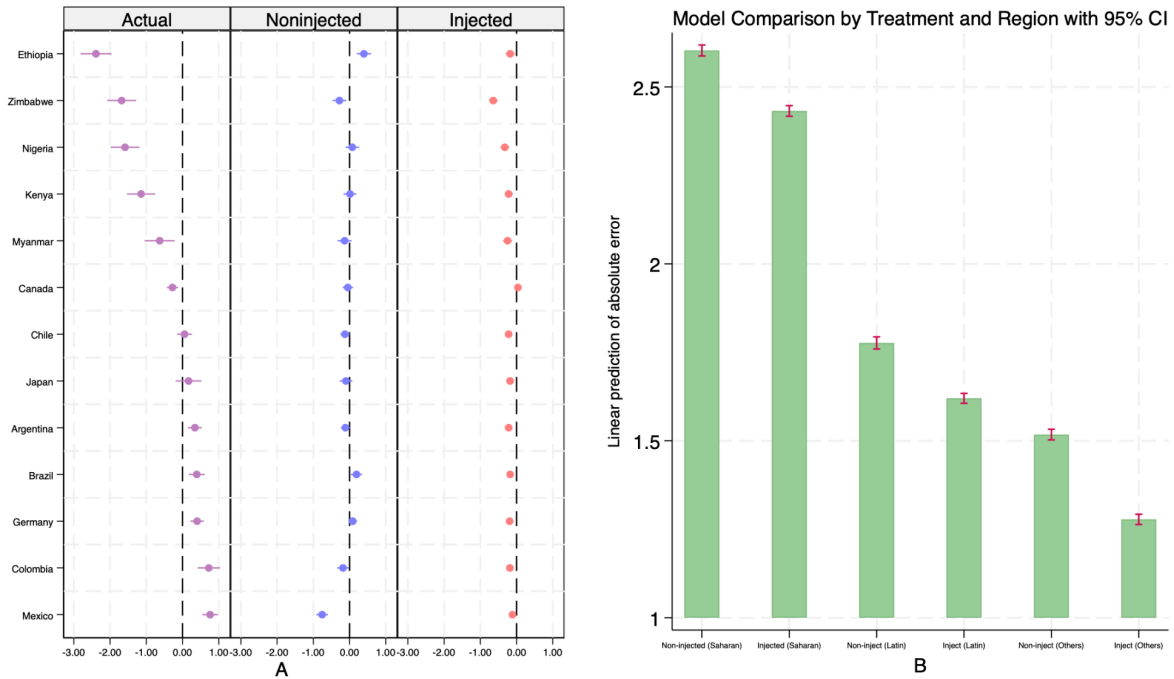

**Figure 7**: Panel A–Coefficient plot of country fixed effects on actual life satisfaction, non-injected LLM predicted life satisfaction, and injected LLM predicted life satisfaction. Reference group: USA. Panel B–Linear prediction of absolute error between actual life satisfaction and LLM predicted life satisfaction by treatment (non-injection vs. injection) and region. 95% confidence intervals are displayed.

**Discussion**

This study examined the effectiveness of LLMs in predicting life satisfaction across different national contexts using large-scale, structured observational data. While the models provided plausible estimates and captured several well-established predictors of well-being, they exhibit systematic variation in performance. Country-specific effects were muted, and key predictors—including income, unemployment, education, and perceived freedom—were often misestimated in both magnitude and significance. These findings suggest that although LLMs offer scalable opportunities for well-being estimation, their utility is constrained by training data biases and limitations in contextual reasoning.

We identified two interrelated sources of error: training data bias and semantic generalisation. Prediction errors were more pronounced in countries with lower internet penetration and GDP per capita, reflecting the influence of digital infrastructure on training exposure. To test semantic generalisation, we used fictional constructs arranged along linguistic gradients. GPT and Gemma exhibited smooth, directional interpolation across unseen but semantically related prompts, even without substantive cues, suggesting a reliance on surface-level similarity rather than conceptual understanding. These effects reversed with prompt polarity and disappeared in placebo conditions. Claude, by contrast, showed minimal extrapolation, reflecting a more conservative generalisation profile.

These mechanisms explain why LLM predictions can appear coherent while systematically deviating from the underlying data. Errors track with digital infrastructure, underscoring the importance of training exposure, while semantic interpolation amplifies dominant discourses at the expense of empirically stronger, but less rhetorically prominent, correlates. This misalignment raises significant risks when LLM outputs are used to inform policy, especially in global development and public health, where misestimated well-being could distort resource allocation, programme evaluation, or public messaging.

To test whether LLM bias can be mitigated, we implemented a targeted injection, introducing a general empirical fact about life satisfaction in Sub-Saharan Africa. This intervention improved Claude's predictions for the targeted region, but also affected outputs for unrelated countries such as those in Latin America. Additionally, the injection altered the model's weighting of individual-level predictors like income and health, suggesting unintended shifts in feature prioritisation. These results emphasise the need for cautious, model-specific calibration of injection strategies to reduce bias without destabilising internal logic.

More broadly, these findings raise concerns about fairness and epistemic reliability. While predictions may appear plausible in well-represented settings, they often rely on linguistic proximity rather than conceptual understanding, which can often lead to superficial coherence. Moreover, LLMs may inherit conceptual biases from dominant but unreplicated research, reinforcing outdated narratives under a veneer of scientific precision. As these models increasingly influence decisions in education, development, and health, it is vital to anchor them in diverse, empirically grounded knowledge. Future development should prioritise improved data representation, robust context injection methods, and evaluation frameworks that reward conceptual fidelity over rhetorical fluency.

**Limitations**

While this study provides insight into LLMs' capacity to estimate life satisfaction across global contexts, several limitations should be acknowledged. First, although we draw on data from 64 countries via the World Values Survey, key subpopulations, such as linguistic minorities and individuals without internet access, are likely underrepresented. This underrepresentation may explain the greater prediction errors observed in low-income and digitally disconnected regions, reinforcing our argument that LLM performance is closely tied to the breadth and inclusivity of training data. Second, the models assessed were not fine-tuned for the task of well-being prediction. Their performance, while broadly reasonable, lags behind traditional statistical models likely due to the absence of task-specific calibration. Third, while our semantic extrapolation experiment reveals how LLMs generalise based on surface-level linguistic similarity, it relies on stylised, synthetic prompts. Further research is needed to assess whether these mechanisms lead to systematic misestimation in real-world policy settings. Lastly, our findings suggest that LLMs may internalise and reproduce associations from widely cited but empirically weak literature. This may contribute to the overemphasis of certain predictors, such as education, despite their relatively modest statistical effects in our analysis.

Addressing these limitations will require more representative training data, task-specific alignment techniques (e.g., domain-focused prompt injection), and curated epistemic inputs that prioritise replicable, high-quality sources. These steps are critical for ensuring that LLMs serve as equitable and trustworthy tools in behavioural and policy research.

**Concluding Remarks**

In summary, current LLMs can capture general patterns in life satisfaction but systematically favour high-resource settings and dominant semantic cues. These limitations are not unique to well-being estimation. Structured self-reports—like life satisfaction—are also increasingly used to infer other human traits, including employment status, health outcomes, and behavioural tendencies. This adaptability underscores both the utility and the vulnerability of LLMs in making human-centred predictions.

If unaddressed, such biases risk marginalising underrepresented populations and entrenching flawed assumptions in real-world applications. Our findings suggest that targeted factual injections can partially correct systematic biases. However, these interventions may also produce unintended spillovers, affecting unrelated regions or predictors. This highlights the need for more refined strategies to guide model behaviour without compromising internal coherence.

As LLMs continue to evolve, progress will depend on integrating more representative training data, refining models' reasoning to align with empirical realities, and developing evaluation frameworks that prioritise conceptual validity over superficial plausibility. These steps are critical to ensure that the deployment of foundation models in human-centred domains is both scientifically rigorous and socially responsible.

## Methods

### Data

We used individual-level data from Wave 7 of the World Values Survey (2017–2022), which provides nationally representative samples for 64 countries. From each country, we randomly selected 1,000 respondents to construct a balanced and harmonised cross-national dataset (N = 64,000). All analyses were pre-registered prior to data analysis (https://aspredicted.org/3xsf-kx78.pdf for the main analysis, https://aspredicted.org/8c6y-2q55.pdf for the semantic similarities experiment, and https://aspredicted.org/n3v6-2rfq.pdf for the targeted injection experiment). All the codes can be found on https://github.com/mitmedialab/wellbeing-LLM/.

### Life Satisfaction Prediction Using Large Language Models

We extracted a comprehensive set of socio-economic, attitudinal, and psychological variables from each respondent, including gender, age group, marital status, education, income decile, health status, perceived freedom, religiosity, employment status, social class, importance of religion and democracy, perceptions of corruption and inequality, trust in institutions, country of residence, and others (see Appendix Table for complete list of covariates). These variables were then used to construct natural language prompts for each individual in the form:

"Based on the following information: Gender: Female, Age: 25–34, Education: Completed secondary school, Employment status: Full-time employed, Health: Good, Perceived freedom: 7/10, … Overall, how satisfied is this person with their life nowadays? Please answer on a 0–10 scale, where zero means not satisfied at all and 10 means completely satisfied."

We submitted these prompts to four state-of-the-art large language models—GPT-4o mini, Claude 3.5 Haiku, LLaMA 3.3 70B, and Gemma 3 27B—via their respective APIs. Each model was prompted independently to return a single numeric life satisfaction score on the 0–10 scale, emulating the WVS item: "All things considered, how satisfied are you with your life as a whole these days?" See Figure 5a for a comparison between the distributions of actual life satisfaction and LLM-predicted values.

**Benchmarking Against Statistical Models**

To benchmark the performance of LLMs, we conducted a standard out-of-sample prediction exercise. The full dataset was randomly split into an 80% training set and a 20% test set. On the training set, we estimated life satisfaction using two standard approaches:

- **Ordinary Least Squares (OLS):** A linear regression of life satisfaction on the complete set of covariates listed in Table 1a of the appendix.
- **Lasso Regression:** A penalised linear model that automatically selects predictive features via L1 regularisation.

These models were then used to generate predicted life satisfaction scores for individuals in the test set. We assessed prediction accuracy using **mean absolute error (MAE)**, which is defined as the absolute difference between actual and predicted life satisfaction scores. Specifically, we computed:

- MAE(OLS) and MAE(Lasso) for statistical models;
- MAE(LLM) for each large language model was calculated by comparing their predicted scores against actual self-reports in the test sample.

This component of the analysis—including the prediction pipeline, model comparisons, and evaluation metrics—was pre-registered at https://aspredicted.org/3xsf-kx78.pdf and is presented in Fig. 1 of the main text.

**Structural Comparison of Predictors and Country Fixed Effects**

To assess whether large language models replicate the structural relationships observed in actual life satisfaction data, we estimated five separate ordinary least squares (OLS) regressions using the full sample of 64,000 individuals:

- One regression with **actual life satisfaction** as the dependent variable, and
- Four regressions with **LLM-predicted life satisfaction** from GPT, Claude, Llama, and Gemma, respectively, as dependent variables.

Each model included the same set of socio-economic and attitudinal covariates listed in Table 1a of the appendix—the use of a standard specification allowed for a direct comparison of coefficient estimates across models. Robust standard errors were computed for all OLS regressions to account for potential heteroskedasticity.

All models also included country fixed effects to account for average differences in life satisfaction across countries, independent of individual-level characteristics. We extracted these fixed effects from each model and compared them across actual and LLM-predicted outcomes to assess how well LLMs capture cross-national variation. Coefficients from these analyses are visualised using STATA's *coefplot* command in Figs. 2 and 3 of the main text.

To trace how national context shapes the *systematic* component of mis-prediction, we proceeded in two steps. First, for each LLM we regressed the individual-level absolute prediction error on the same covariates used in the main OLS models, obtaining fitted values that capture the portion of the error statistically attributable to those characteristics. These fitted values were then averaged within each of the 64 countries to yield a country-level measure of "explained error". Second, we regressed this measure on four macro-economic indicators—the Human Development Index, log GDP per capita, internet-user share, and the Gini coefficient. All macroeconomic data were drawn from publicly available sources corresponding to the

2017–2022 survey period. These indicators were selected to capture key dimensions likely to influence LLM performance across countries: economic development (HDI, GDP), inequality (Gini), and digital exposure (internet penetration)—the latter reflecting the likelihood that national contexts are represented in LLM training data. These relationships were visualised using two-way scatter plots with fitted linear trend lines, as shown in Fig. 5 of the main text.

We also took the country fixed effects, which represent the "unexplained" country-specific effects (e.g., cultural, language, social norm) from the absolute prediction error regressions described above (Fig. 4) and plotted them against four macro-level indicators. These associations are depicted in Fig. 4a of the main text using two-way scatter plots overlaid with linear trend lines.

**Testing LLM Responses to Novel Policy-Like Inputs**

To investigate whether large language models (LLMs) rely on semantic similarity rather than reasoning from first principles when interpreting unfamiliar information, we conducted a four-arm randomised experiment using synthetic variables and fabricated research papers.

We created four experimental scenarios, each involving a pair of synthetic variables that had no actual relationship to life satisfaction. One variable in each pair was arbitrarily assigned a positive association, and the other a negative one. For each scenario, we generated multiple AI-written research papers that presented fabricated empirical evidence supporting these associations with a coefficient of 0.3 SD or -0.3 SD in the case of positive and negative associations, respectively. These papers emulated the structure and tone of peer-reviewed academic publications and were injected as a context prompt for each model. The artificial papers were generated using OpenAI o3.

To evaluate the effects of injection, we used prompts based on individual profiles from the World Values Survey (WVS). Each prompt included a clause referencing one of the synthetic variables (e.g., "...and has a dinosaur as a companion…"), forming a semantic continuum in terms of the cosine similarity between the embedding of each statement using OpenAI text-embedding-3-large, from the positively framed variable (condition 1) to the negatively framed one (condition 7). The experiment included 4 scenarios × 7 conditions × 300 unique profiles per condition, resulting in 8,400 prompts per model.

The four experimental groups were:

1. **Control** – a baseline LLM model with no exposure to fabricated research.
2. **Placebo** – GPT injected with 3 papers involving unrelated synthetic variables never referenced in any prompt, serving as an additional control to account for the general effects of context injection on academic-sounding content.
3. **Original Injection** – LLM injected with 3 placebo papers, as in the second group, and papers linking condition 1 to higher life satisfaction and condition 7 to lower life satisfaction.
4. **Reverse Injection** – LLM injected with the same papers as above, but with the direction of the associations reversed.

The four experimental scenarios, each consisting of a pair of variables and 5 in-between statements, are:

1. Scenario 1
    a. (Variable 1) Use origami birds as messengers
    b. Send paper animals as living mail
    c. Arrange talking paper mobiles
    d. Interact with moving paper sculptures
    e. Listen to decorative ceramic that moves
    f. Speak to house plants
    g. (Variable 2) Speak with household objects
2. Scenario 2
    a. (Variable 1) Wear only clothes made of moondust
    b. Dress in clothing made from stardust particles
    c. Wear garments that shift with emotions
    d. Consume drinks that make you glitter
    e. Dine on fruits that pulse with energy
    f. Taste meals that swirl with new flavors

g. (Variable 1) Often eat food that changes color
    3. Scenario 3
        a. (Variable 1) Drive a laughing car
        b. Ride a bicycle that tell funny stories
        c. Ride equipment that makes melodies
        d. Own garden tools that hum tunes
        e. Tend flower beds that whistle music
        f. Raise a plant that sing rock music
        g. (Variable 2) Adopt a family of singing mushrooms
    4. Scenario 4
        a. (Variable 1) Listen to unicorn voices daily
        b. Hear messages from mythical beings
        c. Detect whispers from fairy friends
        d. Sense the secrets of legendary beings
        e. Befriend creatures that speak riddles
        f. Keep an ancient dragon as a pet
        g. (Variable 2) Have a dinosaur companion

Only the two endpoints (conditions 1 and 7) were explicitly represented during injection. The intermediate conditions (2–6) were never seen by the model and thus serve as untrained transitions. Under the null hypothesis, the model should be indifferent to these intermediate values, and predicted life satisfaction should remain statistically flat across them. See Figure 8a for the cosine similarity matrix for seven statements in each experimental scenario.

We compared predicted life satisfaction across models using the same set of prompts and used OLS to estimate treatment–control differences at each condition level via the following regression equation:

$$LS_i = \alpha + \beta_1 Treatment_i + \beta_2 Condition_i + \beta_3 (Treatment_i \times Condition_i) + \gamma + \varepsilon_i,$$

where $LS_i$ is LLM-predicted life satisfaction for individual $i$, $Treatment_i$ is a binary indicator for assignment to the treatment group, $Condition_i$ is a categorical variable representing the semantic condition level. The model includes intervention fixed effects, γ, and $\varepsilon_i$ is the error term. Our primary parameter of interest is $\beta_3$, which captures the treatment–control difference at each condition level. Standard errors are clustered at the individual level to account for repeated observations.

If significant effects appear only at conditions 1 and 7, this would suggest that injection influenced predictions only where trained, consistent with proper localisation. In contrast, systematic effects across untrained intermediate conditions would indicate semantic generalisation—that is, the model is extrapolating from the training signal based on the surface meaning of the inserted clause. See Fig 4a for the illustration of the null hypothesis.

The reversed injection condition tests whether this extrapolation is directionally flexible. The placebo condition functions as a second control group, helping to disentangle the effects of content-specific training from those driven by generic exposure to academic-sounding text. Together, the four experimental arms allow us to isolate the causal impact of content-specific injection on semantic generalisation in LLMs.

**Targeted Injection Experiment**

To explore a potential method for mitigating systematic biases in LLM predictions, particularly those that affect underrepresented countries, we conducted a two-arm randomized controlled experiment. Our focus was on injecting broad contextual information into the model. Given Claude's tendency to overestimate life satisfaction in Sub-Saharan African countries, we aimed to determine whether this method (i.e., injecting the statement "Sub-Saharan African countries tend to report lower average life satisfaction compared to many other regions of the world") could reduce the overestimations in these countries without impacting predictions for others.

We tested the method on three country groups: 5 accurately represented countries (Canada, Germany, Japan, Myanmar, and the United States), 4 Sub-Saharan countries (Ethiopia, Zimbabwe, Nigeria, and Kenya), and 5 Latin American countries (Argentina, Brazil, Chile, Colombia, and Mexico), with 1,000 unique profiles per country, resulting in 15,000 observations. The results were compared with the actual reported life satisfaction and LLM prediction without information injection. We then re-estimated the full life satisfaction specification and reported the country fixed effects separately by treatment group. Additionally, we re-ran the regression of absolute life satisfaction error to obtain its linear predictions, which we then compared across treatments.

# Appendix

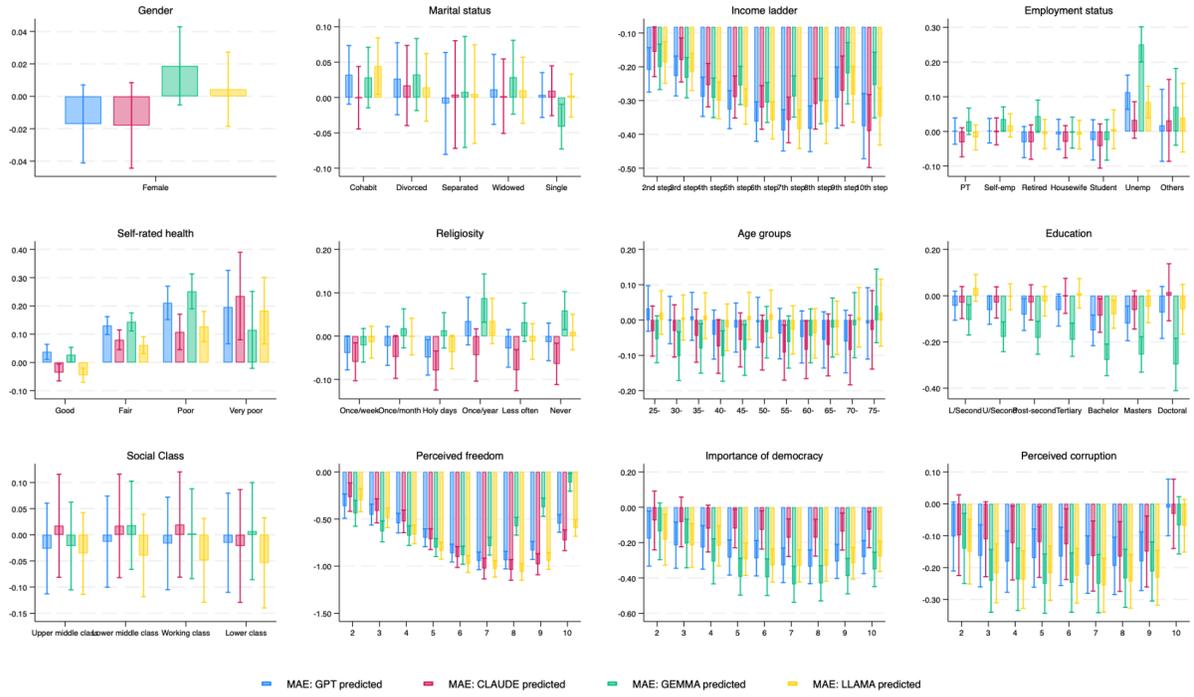

**Fig 1a:** Coefficient plot of covariates predicting the absolute difference between actual and LLM-predicted life satisfaction

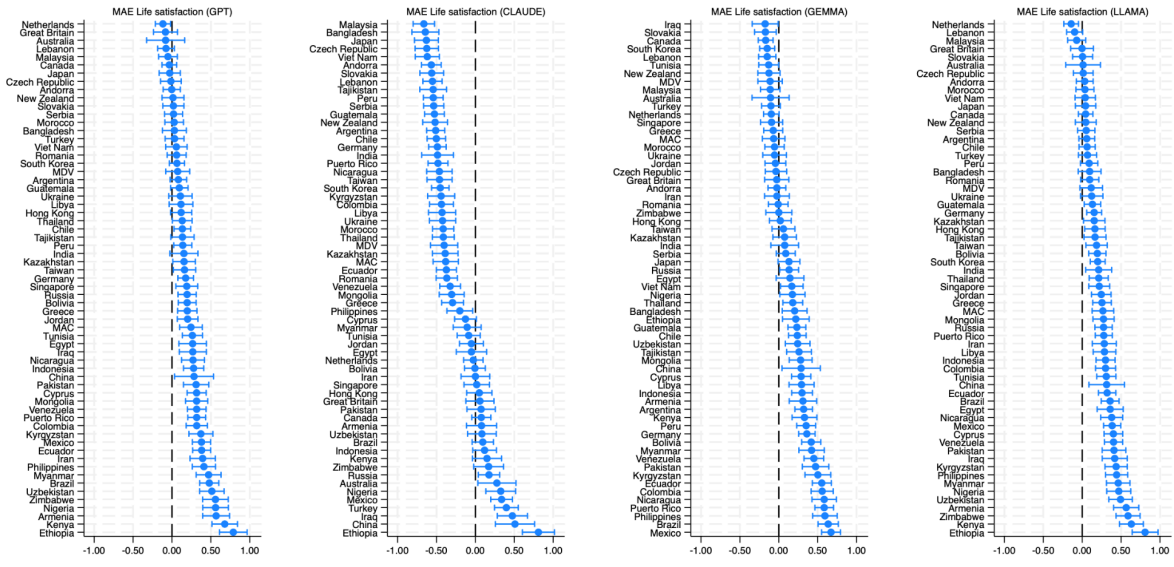

**Fig 2a:** Coefficient plot of country fixed effects from the absolute difference between actual and LLM-predicted life satisfaction.

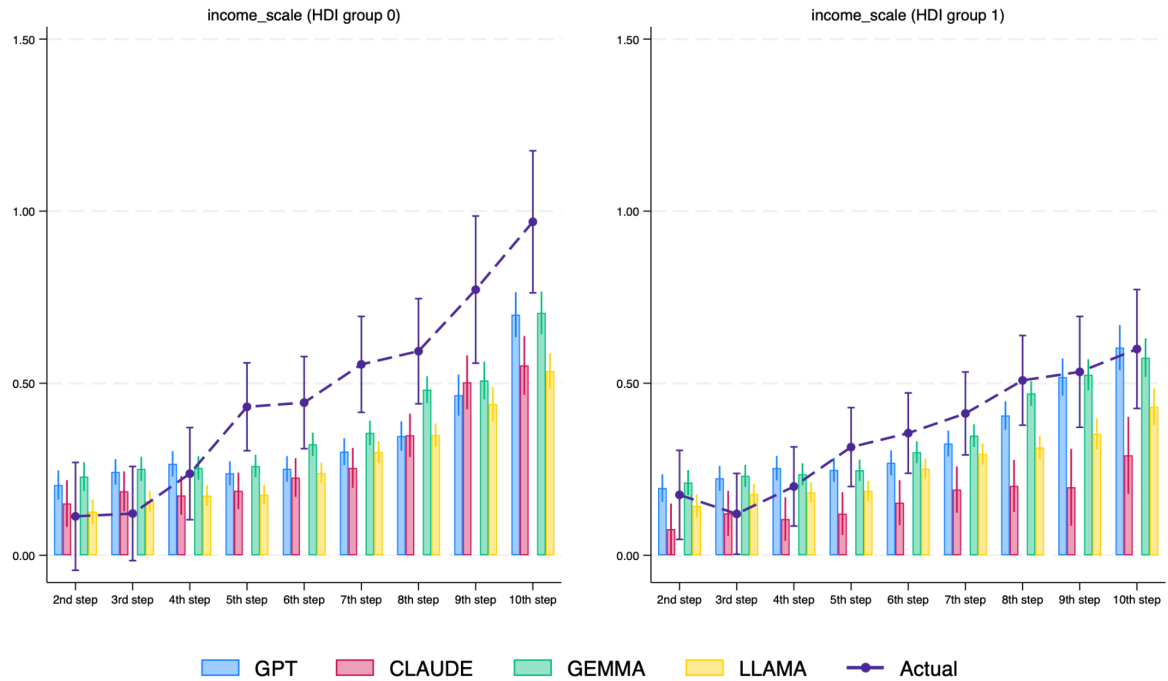

**Figure 3a:** The estimated relationship between income decile and life satisfaction across countries with below-median (left) and above-median (right) Human Development Index (HDI). The dotted lines represent the observed (actual) income–life satisfaction relationships with 95% confidence intervals, while the bars show predicted values from four large language models (GPT, Claude, Gemma, and LLaMA). Although the actual gradient is steeper in lower-HDI countries, all models estimate similar slopes across both groups—failing to capture the stronger role of income in less developed contexts.

**Descriptive statistics**

| Table 1a: Summary Statistics | | | | | |
|---|---|---|---|---|---|
| **Dependent variables** | **N** | **Mean** | **SD** | **Min** | **Max** |
| Life satisfaction (actual) | 63647 | 7.008 | 2.281 | 1 | 10 |
| Life satisfaction (GPT) | 64000 | 6.932 | 1.273 | 1 | 10 |
| Life satisfaction (Claude) | 64000 | 7.338 | 1.356 | 1 | 10 |
| Life satisfaction (Gemma) | 64998 | 6.004 | 1.192 | 2 | 10 |
| Life satisfaction (Llama) | 64998 | 6.953 | 1.071 | 1 | 10 |
| **Covariates** | **N** | **Mean** | **SD** | **Min** | **Max** |
| Importance: Family (Q1) | 63894 | 1.104 | 0.349 | 1 | 4 |
| Importance: Friends (Q2) | 63779 | 1.712 | 0.746 | 1 | 4 |
| Importance: Leisure time (Q3) | 63641 | 1.796 | 0.793 | 1 | 4 |
| Importance: Politics (Q4) | 63266 | 2.629 | 0.989 | 1 | 4 |
| Importance: Work (Q5) | 63248 | 1.529 | 0.770 | 1 | 4 |
| Importance: Religion (Q6) | 63378 | 1.907 | 1.059 | 1 | 4 |
| Gender: Female (Q260) | 63946 | 1.526 | 0.499 | 1 | 2 |
| Age categories: 5-year intervals | 63998 | 5.236 | 3.420 | 0 | 16 |
| Self-reported health (Q47) | 63827 | 2.189 | 0.880 | 1 | 5 |
| Perceived freedom (Q48) | 63409 | 7.187 | 2.291 | 1 | 10 |
| Gone without food(Q51) | 63654 | 3.484 | 0.856 | 1 | 4 |
| Felt unsafe from crime (Q52) | 63588 | 3.443 | 0.879 | 1 | 4 |
| Gone without medicine (Q53) | 63577 | 3.348 | 0.932 | 1 | 4 |
| Gone without a cash income (Q54) | 63587 | 3.156 | 1.014 | 1 | 4 |
| Gone without a safe shelter (Q55 | 63470 | 3.718 | 0.702 | 1 | 4 |
| Standard of living (Q56) | 62825 | 1.737 | 0.864 | 1 | 3 |
| Most people can be trusted (Q57) | 63070 | 1.781 | 0.413 | 1 | 2 |
| Confidence: The Press (Q66) | 62446 | 2.714 | 0.876 | 1 | 4 |
| Confidence: The Police (Q69) | 62158 | 2.375 | 0.947 | 1 | 4 |
| Confidence: The Government (Q71) | 61628 | 2.664 | 0.987 | 1 | 4 |
| Confidence: Politics (Q72) | 61638 | 3.006 | 0.893 | 1 | 4 |
| Confidence: Universities (Q75) | 61305 | 2.170 | 0.858 | 1 | 4 |
| Confidence: Election (Q76) | 61174 | 2.642 | 0.958 | 1 | 4 |
| Pref. income inequality (Q106) | 63094 | 6.349 | 3.008 | 1 | 10 |
| Perceptions of corruption (Q112) | 62913 | 7.702 | 2.432 | 1 | 10 |
| Perceptions of security (Q131) | 63562 | 2.004 | 0.825 | 1 | 4 |
| How important is God? (Q164) | 62285 | 7.621 | 3.082 | 1 | 10 |
| Religious service (Q171) | 63253 | 4.072 | 2.179 | 1 | 7 |
| Are you religious? (Q173) | 62174 | 1.443 | 0.637 | 1 | 3 |
| How important is democracy? (Q25 | 62856 | 8.352 | 2.149 | 1 | 10 |
| Nation pride (Q254) | 63133 | 1.593 | 0.863 | 1 | 5 |
| Immigrant status (Q263) | 63789 | 1.058 | 0.234 | 1 | 2 |
| Citizen (Q269) | 61778 | 1.024 | 0.152 | 1 | 2 |
| Household size (Q270) | 63362 | 4.009 | 2.281 | 1 | 63 |
| Marital status (Q273) | 63608 | 2.678 | 2.157 | 1 | 6 |
| Number of children (Q274) | 61775 | 1.799 | 1.743 | 0 | 22 |
| Highest education (Q275) | 63366 | 3.535 | 2.015 | 0 | 8 |
| Employment status (Q279) | 63264 | 3.141 | 2.063 | 1 | 8 |
| Occupation (Q281) | 59849 | 4.011 | 3.033 | 0 | 11 |
| Working in the public sector? (Q284) | 47920 | 1.821 | 0.562 | 1 | 3 |
| Chief wage earner (Q285) | 62107 | 1.538 | 0.499 | 1 | 2 |
| Social class (Q287) | 61378 | 3.238 | 0.975 | 1 | 5 |
| Income scale (Q288) | 62171 | 4.913 | 2.071 | 1 | 10 |
| Religion (Q289) | 63121 | 2.993 | 2.542 | 0 | 9 |
| Literacy | 30904 | 1.091 | 0.287 | 1 | 2 |
| Settlement: Urban/rural | 63855 | 3.059 | 1.509 | 1 | 5 |

**Note:** We included missing dummies for all variables in the regressions.

**1. Personal Values & Life Priorities**

1. Family importance (Q1)
2. Friends importance (Q2)
3. Leisure importance (Q3)
4. Work importance (Q5)
5. Social trust (Q57)

**2. Religious & Spiritual Beliefs**

1. Religion importance (Q6)
2. God importance (Q164)
3. Religious service (Q171)
4. Religious identity (Q173)
5. Religion type (Q289)

**3. Political Trust & Civic Engagement**

1. Politics importance (Q4)
2. Press confidence (Q66)
3. Police confidence (Q69)
4. Government confidence (Q71)
5. Politics confidence (Q72)
6. Universities confidence (Q75)
7. Election confidence (Q76)
8. Corruption perceptions (Q112)
9. Democracy importance (Q252)
10. National pride (Q254)

**4. Economic Security & Social Status**

1. Income deprivation (Q54)
2. Living standard (Q56)
3. Income inequality preference (Q106)
4. Employment status (Q279)
5. Occupation (Q281)
6. Public sector work (Q284)
7. Chief wage earner (Q285)
8. Social class (Q287)
9. Income scale (Q288)

**5. Health & Personal Well-being**

1. Self-reported health (Q47)
2. Perceived freedom (Q48)
3. Crime safety (Q52)
4. Security perceptions (Q131)
5. Food deprivation (Q51)
6. Medicine deprivation (Q53)
7. Shelter deprivation (Q55)

**6. Demographic & Social Background**

1. Gender (Q260)
2. Age categories
3. Immigrant status (Q263)
4. Citizenship (Q269)

5. Household size (Q270)
6. Marital status (Q273)
7. Number of children (Q274)
8. Education level (Q275)
9. Literacy
10. Urban/rural (Settlement)

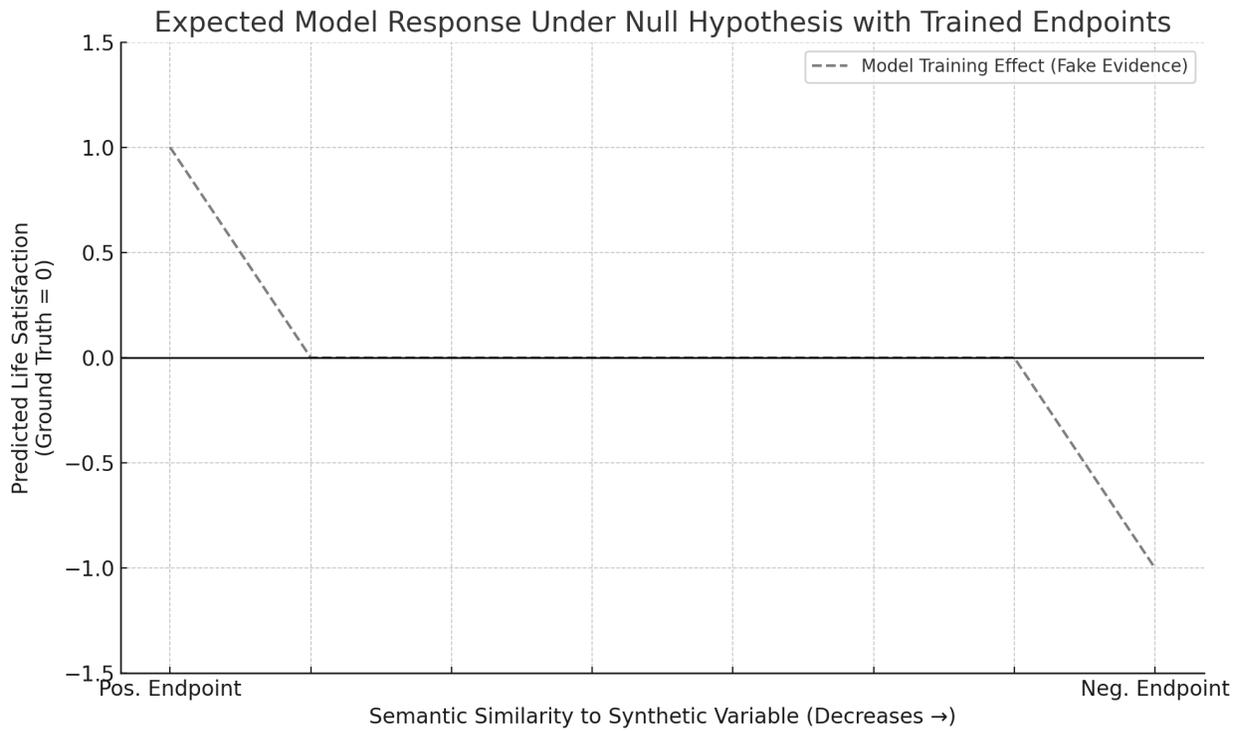

**Fig 4a: Expected pattern of LLM-predicted life satisfaction across a semantic continuum.** Under the null hypothesis, predicted life satisfaction should remain at zero for all transitional prompts. Due to context injection on synthetic papers, only the two endpoints were trained to have positive or negative associations with life satisfaction, producing deviations at the extremes despite a true underlying relationship of zero throughout.

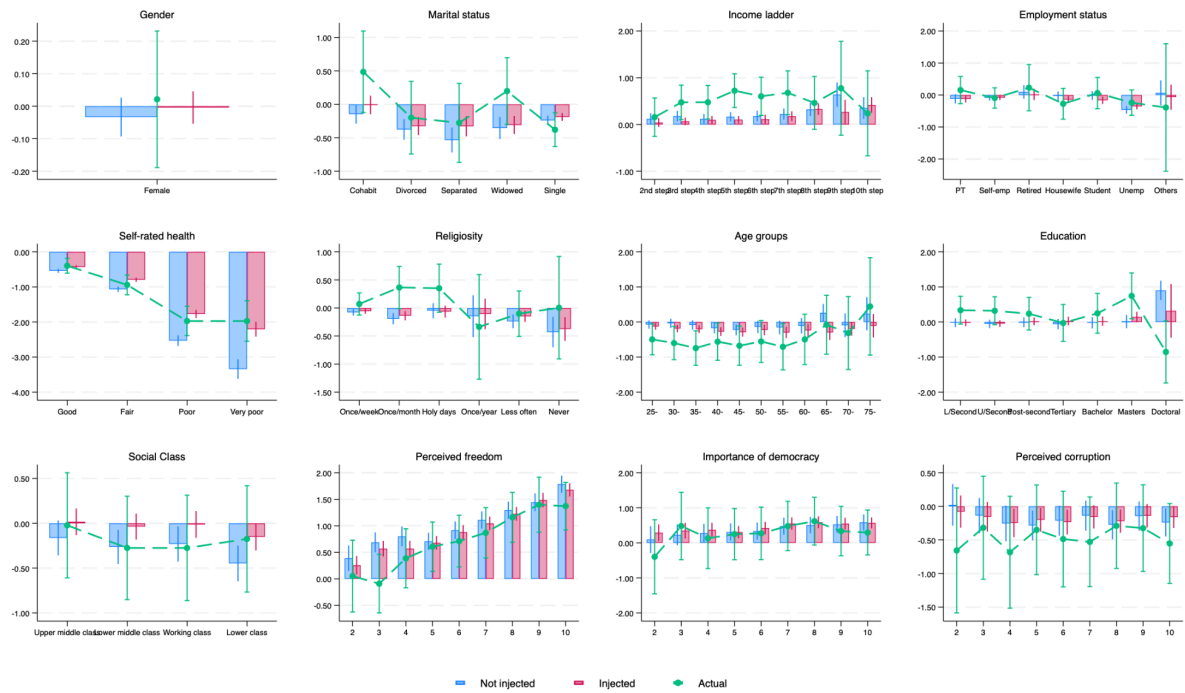

**Fig 5a:** Coefficient plot of socio-economic and attitudinal predictors of actual, non-injected, and injected LLM-predicted life satisfaction for Sub-Saharan countries: Ethiopia, Zimbabwe, Nigeria, and Kenya. Reference groups are: male (gender), married (marital status), first step (income ladder), full-time employed (employment status), very good (self-rated health), more than once a week (religiosity), younger than 25 (age groups), no formal education (education), upper class (social class), 1.little perceived freedom (perceived freedom), 1.democracy is not important (importance of democracy), and 1.zero corruption (perceived corruption). 95% confidence intervals are displayed.

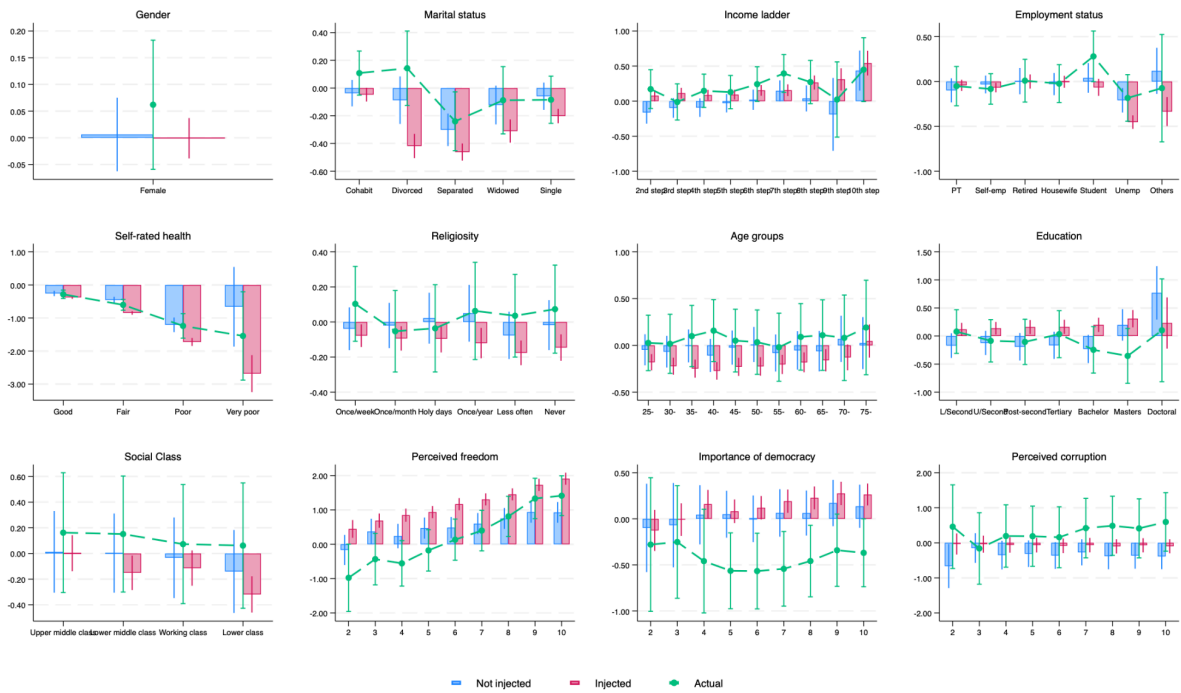

**Fig 6a:** Coefficient plot of socio-economic and attitudinal predictors of actual, non-injected, and injected LLM-predicted life satisfaction for Latin American countries: Argentina, Brazil, Chile, Colombia, and Mexico. Reference groups are: male (gender), married (marital status), first step (income ladder), full-time employed (employment status), very good (self-rated health), more than once a week (religiosity), younger than 25 (age groups), no formal education (education), upper class (social class), 1.little perceived freedom (perceived freedom), 1.democracy is not important (importance of democracy), and 1.zero corruption (perceived corruption). 95% confidence intervals are displayed.

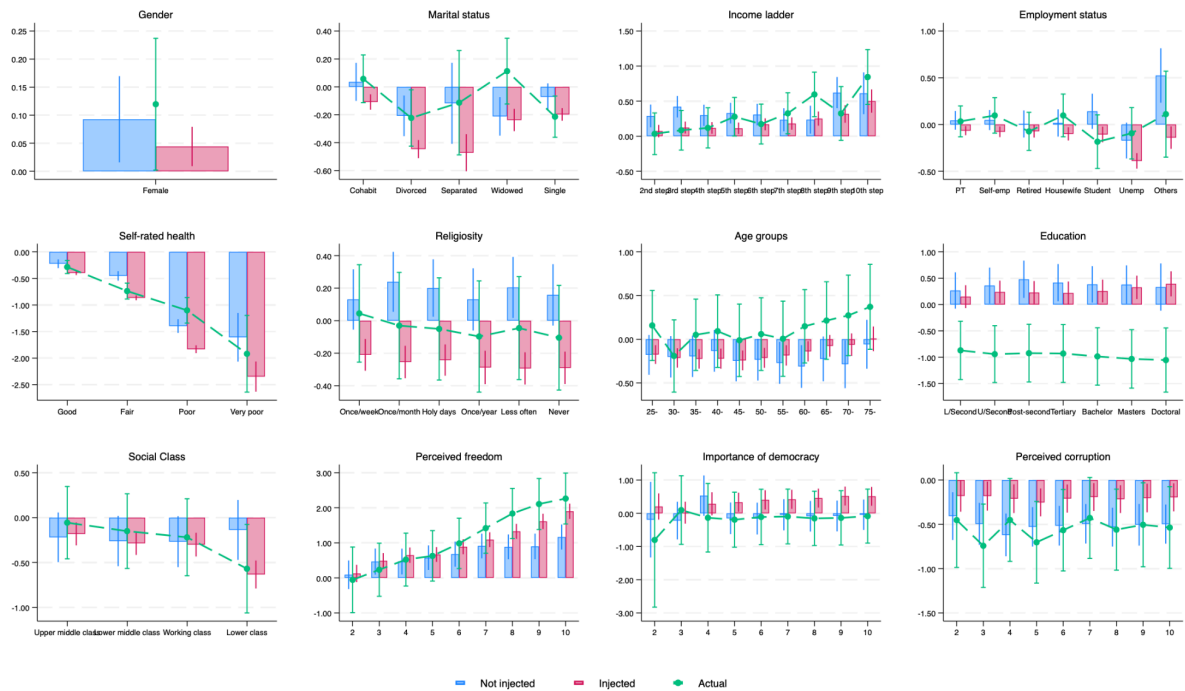

**Fig 7a:** Coefficient plot of socio-economic and attitudinal predictors of actual, non-injected, and injected LLM-predicted life satisfaction for 'Others' countries: Canada, Germany, Japan, Myanmar, and the United States. Reference groups are: male (gender), married (marital status), first step (income ladder), full-time employed (employment status), very good (self-rated health), more than once a week (religiosity), younger than 25 (age groups), no formal education (education), upper class (social class), 1.little perceived freedom (perceived freedom), 1.democracy is not important (importance of democracy), and 1.zero corruption (perceived corruption). 95% confidence intervals are displayed.

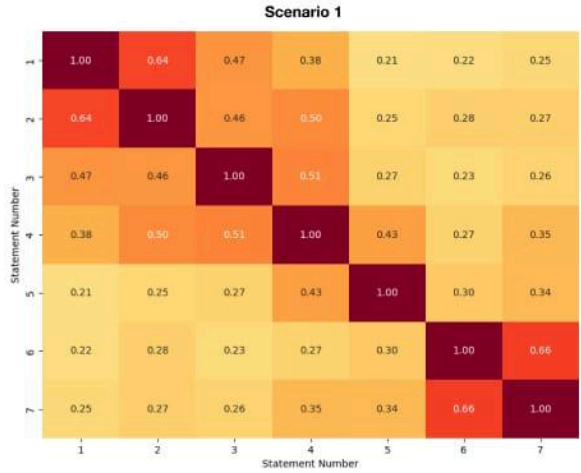
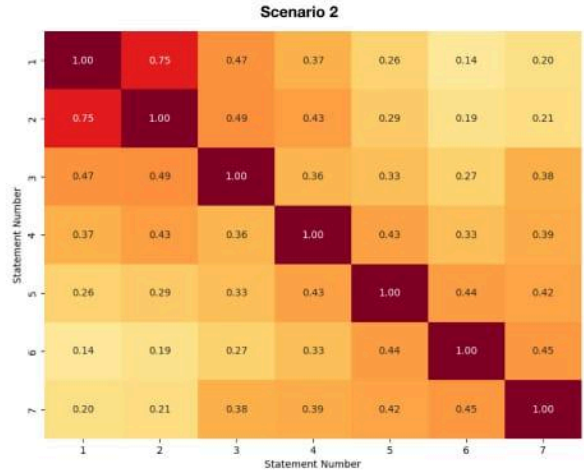
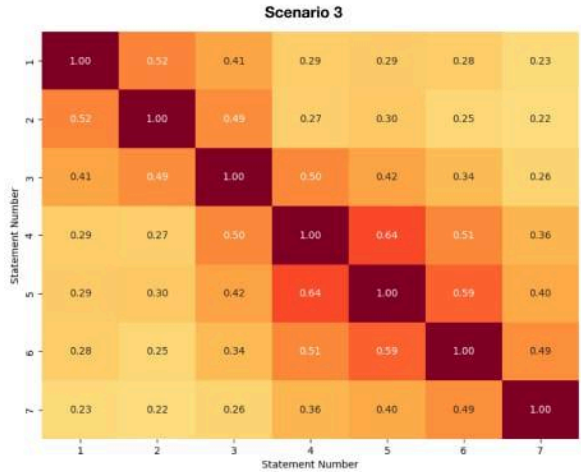
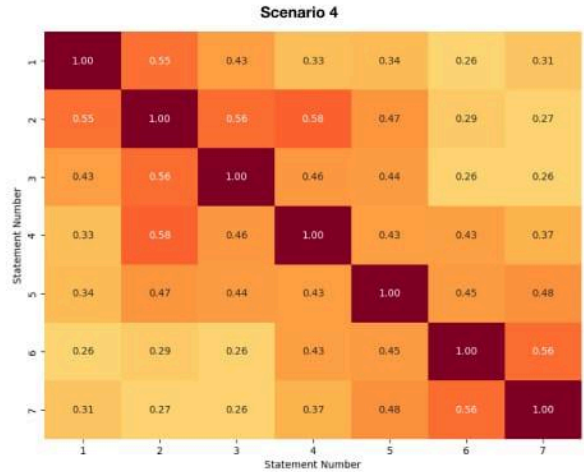

**Figure 8a**: Cosine similarity matrix for seven statements in each experimental scenario based on OpenAI text-embedding-3-large. The statements are designed to form a semantic continuum between the two endpoints (statement 1 and statement 7).